\documentclass[12pt,preprint]{aastex}

\newcommand{\teff}{$T_{\rm eff}$} 
\newcommand{\kms}{km s$^{-1}$}
\begin{document}

\title{Chemical abundances in giants stars of the tidally disrupted 
globular cluster NGC 6712 from high-resolution infrared 
spectroscopy}

\author{David Yong}
\affil{Research School of Astronomy and Astrophysics, Australian National 
University, Mount Stromlo Observatory, Cotter Road, Weston Creek, ACT 2611, 
Australia}
\email{yong@mso.anu.edu.au}

\author{Jorge Mel{\' e}ndez}
\affil{Centro de Astrof{\' i}sica da Universidade do Porto, Rua das Estrelas, 
4150-762 Porto, Portugal}
\email{jorge@astro.up.pt}

\author{Katia Cunha\altaffilmark{1}}
\affil{National Optical Astronomy Observatory, Casilla 603, La Serena, Chile} 
\email{kcunha@noao.edu}

\author{Amanda I.\ Karakas}
\affil{Research School of Astronomy and Astrophysics, Australian National 
University, Mount Stromlo Observatory, Cotter Road, Weston Creek, ACT 2611, 
Australia}
\email{akarakas@mso.anu.edu.au}

\author{John E.\ Norris}
\affil{Research School of Astronomy and Astrophysics, Australian National 
University, Mount Stromlo Observatory, Cotter Road, Weston Creek, ACT 2611, 
Australia}
\email{jen@mso.anu.edu.au}

\author{Verne V.\ Smith}
\affil{National Optical Astronomy Observatory, Casilla 603, La Serena, Chile}
\email{vsmith@noao.edu}

\altaffiltext{1}{On leave from
Observat\'orio Nacional; Rio de Janeiro, Brazil}

\begin{abstract}
We present abundances of C, N, O, F, Na, and Fe in 
six giant stars of the tidally disrupted globular cluster NGC 6712. 
The abundances were derived by comparing synthetic spectra with 
high resolution infrared spectra obtained with the Phoenix 
spectrograph on the Gemini South telescope. We find large star-to-star 
abundance variations of the elements C, N, O, F, and Na. NGC 6712 
and M4 are the only globular clusters in which F has been measured in 
more than two stars, and both clusters reveal F abundance variations 
whose amplitude is comparable to, or exceeds, that of O, 
a pattern which may be produced in $M$ $\gtrsim$ 5$M_\odot$ 
AGB stars. 
Within the limited samples, 
the F abundance in globular clusters is lower than in 
field and bulge stars at the same metallicity. 
NGC 6712 and Pal 5 are tidally disrupted globular 
clusters whose red giant members exhibit O and Na abundance variations 
not seen in comparable metallicity field stars. Therefore, globular clusters 
like NGC 6712 and Pal 5 cannot contribute many field stars and/or 
field stars do not form in environments with chemical enrichment 
histories like that of NGC 6712 and Pal 5. Although our sample size 
is small, from the amplitude of the O and Na abundance 
variations, we infer a large 
initial cluster mass and tentatively confirm that NGC 6712 was once 
one of the most massive globular clusters in our Galaxy. 
\end{abstract}


\section{Introduction}
\label{sec:intro}

The formation and evolution of our Galaxy remains one of the great 
unanswered questions in modern astronomy. 
\citet{els62} suggested formation via the monolithic collapse of a 
gaseous protocloud on a timescale of 10$^8$ years. 
\citet{sz78} challenged this notion by proposing that 
the halo formed through the accretion of independent fragments 
over a longer period, 10$^9$ years. 
These seminal works studied Galactic archaeology using 
the kinematics and metallicities of stars and globular clusters 
in the disk and halo. Today, Galaxy formation is discussed within 
the context of $\Lambda$CDM cosmology and hierarchical structure 
formation \citep{white78,freeman02} with the ongoing accretion of the 
Sagittarius dwarf galaxy being the most prominent example 
\citep{ibata94}. 

Another 
mechanism for populating the disk and halo is through the destruction 
of globular clusters via tidal shocks, two body relaxation etc 
\citep{gnedin97}. 
Although the current mass in globular clusters is small, 
the initial globular cluster population may have been 
considerably larger than the present population \citep{gnedin97}. 
The globular cluster Palomar 5 exhibits large tidal tails that extend 
over 10 degrees 
and contain more mass than the remaining cluster 
\citep{odenkirchen01,odenkirchen03}. 
Therefore, Pal 5 is in the process of being tidally disrupted and 
is currently contributing stars to the disk and halo. 

Chemical abundances place strong constraints upon the fraction of halo 
and disk stars that may come from disrupted globular clusters and/or 
the types of globular clusters that may populate the disk and halo. 
Specifically, every well studied Galactic globular cluster exhibits large 
star-to-star abundance variations for the light elements from C to Al 
\citep{smith87,kraft94,gratton04}. Although the amplitude may vary 
from cluster to cluster, the abundances of C and O are low when N is 
high, O and Na are anticorrelated as are Mg and Al. 
Indeed, Str{\"o}mgren photometry reveals that every globular cluster has 
large star-to-star variations in the $c_1$ = $(u-v)-(v-b)$ index at all 
evolutionary stages \citep{grundahl00}. The Str{\"o}mgren $u$ filter includes 
the 3360\AA\ NH molecular lines, and \citet{6752nh} recently showed 
that the N abundances are directly correlated with the $c_1$ index. 
Therefore, it is likely that 
all globular clusters possess large N abundance variations at all 
evolutionary stages. 
Although hydrogen burning at 
high temperatures may explain the observed abundance patterns 
\citep{langer93,langer95,denissenkov98,karakas03}, the source of the 
nucleosynthesis and the nature of the 
pollution mechanism remain unknown. 
Intermediate-mass ($\sim$3 to 8$M_{\odot}$) asymptotic giant branch 
(AGB) stars were the assumed polluters owing to the mono-metallic nature 
of most GCs, even though detailed AGB models have so far mostly failed 
to match the observations \citep{fenner04,karakas06b}. 
Nevertheless, 
these abundance patterns seen in every cluster have rarely, if ever, been 
observed in field stars to date \citep{pilachowski96,gratton00b}. 

\citet{smith02} conducted a detailed abundance analysis of four bright 
giant stars in Pal 5 and found variations of O, Na, and Al. 
(No abundance measurements have been performed upon stars in the 
tidal tails of Pal 5.) 
While 
most stars lost from a tidally disrupted cluster would be main sequence
stars, abundance variations of O, Na, and Al have now been identified on the 
main sequences of globular clusters \citep{gratton01,cohen05}. 
Since no radial gradients are associated with the O to Al abundance 
variations 
(with the exception of 47 Tucanae [\citealt{nf79,briley97}]), 
observations of red giants in the cluster should be 
equivalent to observing red giants in the tidal tails. That 
abundance variations of O, Na, and Al are found in Pal 5 
suggests that clusters like 
Pal 5 cannot provide many field stars and/or field stars do not form 
in environments with chemical enrichment histories similar to Pal 5. 
Of great interest for our understanding of Galactic and globular cluster 
formation would be the identification of 
clusters undergoing tidal disruption in which no light element 
abundance variations are detected. 

Of the large sample of globular clusters studied by \citet{paresce00} 
using the Hubble Space Telescope, all have mass functions (as inferred 
from their luminosity functions) which peak at 0.25M$_\odot$. 
Not surprisingly, the mass function of Pal 5 
is flatter than other clusters revealing significant depletions of 
low mass stars presumably stripped by the Galactic tidal field 
\citep{koch04}. The globular cluster NGC 6712 is a small and sparse 
globular cluster whose mass function peaks at 0.75M$_\odot$ 
instead of 0.25M$_\odot$ \citep{demarchi99,andreuzzi01}. 
That is, NGC 6712 is the only cluster 
whose mass function decreases with decreasing mass. With an 
orbit penetrating deep into the bulge, $R_{\rm pericentric}$ = 0.9 kpc 
\citep{dinescu99}, tidal forces have stripped away a substantial 
fraction of NGC 6712's lower mass stellar population. 
Calculations suggest that NGC 6712 may have lost up to 99\% of its original 
mass \citep{takahashi00}. 
All that remains of NGC 6712 is a remnant core of a cluster that 
was probably once one of the most massive in the Galaxy. 
The presence of a high luminosity x-ray source and a surprisingly large 
blue straggler population reinforce the idea that NGC 6712 was once 
much more massive and concentrated \citep{paltrinieri01}. 
Therefore, NGC 6712 has almost 
certainly contributed stars to the disk and/or halo. 
Previous abundance analyses of NGC 6712 only 
considered one post-AGB star \citep{jasniewicz04,mooney04} 
whose composition may not reflect 
the composition of the cluster due to the rich nucleosynthesis occurring 
in the late phases of stellar evolution.
In this paper, we present the first detailed  
chemical abundance analysis of bright red giant stars 
in this tidally disrupted globular cluster. 

\section{Observations, data reduction, and analysis}
\label{sec:data}

\begin{figure}
\epsscale{0.8}
\plotone{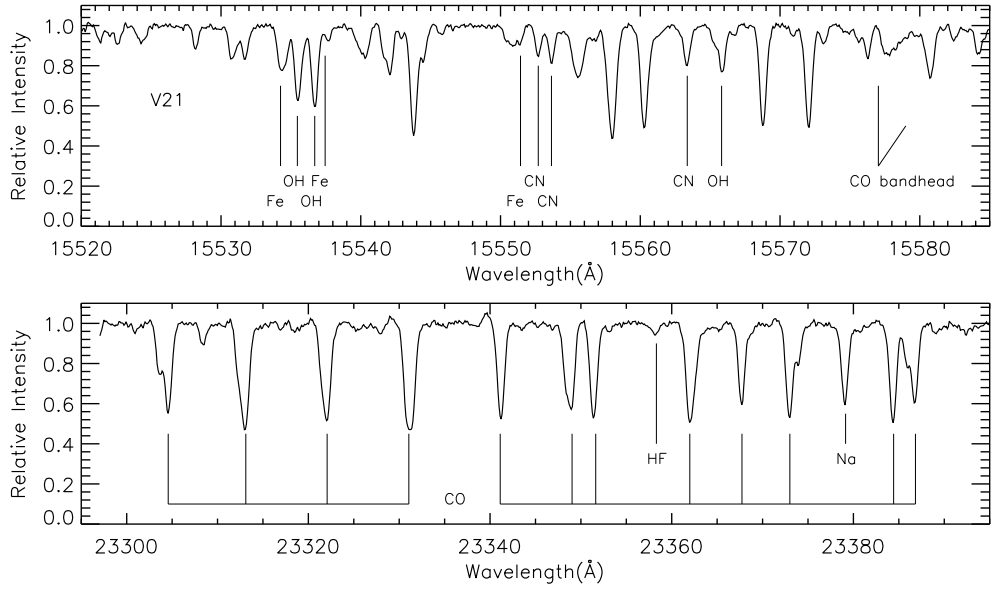}
\caption{Spectra of V21 for the two wavelength regions. 
Lines used in the abundance analysis are indicated. 
\label{fig:spectra}}
\end{figure}

NGC 6712 lies in the direction of the Galactic bulge in a region of 
high visual extinction. While the brightest giants are relatively faint at 
visual wavelengths ($V$ $\simeq$ 13.5), these giants are very bright 
at infrared wavelengths ($H$ $\simeq$ $K$ $\simeq$ 8.3). 
\citet{cudworth88} measured proper motions from which membership 
probabilities were determined. For bright giants with 
membership probabilities $>$ 90\%, optical ($V$ vs.\ $B-V$) and 
infrared ($K$ vs.\ $J-K$) color-magnitude
diagrams were constructed using the \citet{cudworth88} and 
2MASS \citep{2mass} photometry. The six brightest stars were 
observed using the Gemini South telescope and the Phoenix 
spectrograph \citep{phoenix} in service mode in July and August 2007. 
The program stars are listed in Table \ref{tab:prog} and the log 
of observations is shown in Table \ref{tab:obs}. 
We used the 0.35\arcsec\ slit which provided a spectral resolution 
of $R$ = 50,000. 
All program stars were observed at two positions along the slit 
separated by 5\arcsec\ on the sky through two 
filters: 
the H6420 filter provided wavelength coverage from 15520\AA\ to 15585\AA\ 
and 
the K4308 filter provided wavelength coverage from 23300\AA\ to 23400\AA. 
The exposure times per star 
ranged from 520 seconds for the H-band observation of V10 
to 36 minutes for the K-band observation of LM10. 
The signal-to-noise ratios (S/N) 
exceed 150 per resolution element for each setting in each star. 
For each setting on each night, our observing program included 
a radial velocity standard, a hot star for telluric line removal, 
10 flat field exposures, and 10 dark exposures. 
Wavelength calibrated spectra were produced using 
standard reduction 
procedures for infrared data described by \citet{smith02b} and 
\citet{melendez03} 
with the IRAF\footnote{IRAF (Image Reduction and Analysis
Facility) is distributed by the National Optical Astronomy
Observatory, which is operated by the Association of Universities
for Research in Astronomy, Inc., under contract with the National
Science Foundation.} package of programs. 
Examples of reduced spectra 
are shown in Figure \ref{fig:spectra}. 

Radial velocities were measured by cross correlating the cluster 
spectra against the radial velocity standards. For each star, we 
obtained a radial velocity measure from the H-band and the K-band and 
the velocities measured from each region were in good agreement for a 
given star. In Table \ref{tab:rv}, we report the radial velocities and 
for our six stars we find a mean cluster radial velocity 
$V_{\rm rad}$ = $-$109.0 \kms\ ($\sigma$ = 5.0 \kms) 
which is in good agreement 
with the value in the \citet{harris96} catalog, 
$V_{\rm rad}$ = $-$107.5 \kms\ 
as well as the value measured by 
\citet{jasniewicz04} for their post-AGB star, $V_{\rm rad}$ = $-$116.4 \kms. 

The stellar parameters were derived in the following way. The 
effective temperature, \teff, was calculated using the \citet{ramirez05} 
\teff:color:[Fe/H] calibrations for giant 
stars. We used the ($B-V$), ($V-J$), ($V-H$), and 
($V-K$) colors from the \citet{cudworth88} and 2MASS \citep{2mass} photometry, 
$E(B-V)$ = 0.43 \citep{cudworth88}, and [Fe/H] = $-$1.01 from the 
\citet{harris96} catalog. The final \teff\ was the mean of the individual 
\teff\ values from each color weighted by the uncertainties for each 
color calibration. The surface gravity, $\log g$, was determined using 
\teff, a distance modulus of $(m-M)_V$ = 15.6 \citep{harris96}, 
bolometric corrections BC(V) from \citet{alonso99b}, and assuming a mass 
of 0.8M$_\odot$. The microturbulent velocity was determined using 
the following relation, $\xi_t$ = 4.2 $-$ 6$\times$10$^{-4}$ \teff, 
adopted from the optical analysis 
by \citet{melendez08} of thick disk and bulge stars with 
comparable stellar parameters. The stellar parameters 
are given in Table \ref{tab:param}. 

We found that errors in the distance modulus of $\pm$0.2 led 
to changes of 0.08 dex in $\log g$ and that changes of $\pm$0.02 mag 
in reddening resulted in \teff\ errors of 20K. 
Had we adopted the \citet{alonso99b} \teff:color:[Fe/H] calibration for giant 
stars, our values for \teff\ would be 93K ($\sigma$ = 29K) hotter and $\log g$ 
would be 0.07 dex ($\sigma$ = 0.05 dex) higher. 
We note that the \citet{schlegel98} dust maps give a reddening $E(B-V)$ = 0.39 
and that \citet{paltrinieri01} find a very low value of $E(B-V)$ = 0.33. 
Had we adopted the lowest published value 
$E(B-V)$ = 0.33, our \teff\ would be 108K 
($\sigma$ = 20K) cooler and $\log g$ 
would be 0.03 dex ($\sigma$ = 0.04 dex) lower. 
We estimate that internal uncertainties in the stellar parameters are 
\teff\ $\pm$ 50K, $\log g$ $\pm$ 0.2 dex, and $\xi_t$ $\pm$ 0.2 \kms. 
While the zero-point of our derived 
abundances would shift, the amplitude of the 
star-to-star abundance variation for C, N, O, F, and Na would remain 
similar regardless of the adopted stellar parameters provided they 
were homogeneously applied. Therefore, our 
conclusions do not depend upon the adopted stellar parameters, within 
a reasonable error range. 
 
\begin{figure}
\epsscale{0.8}
\plotone{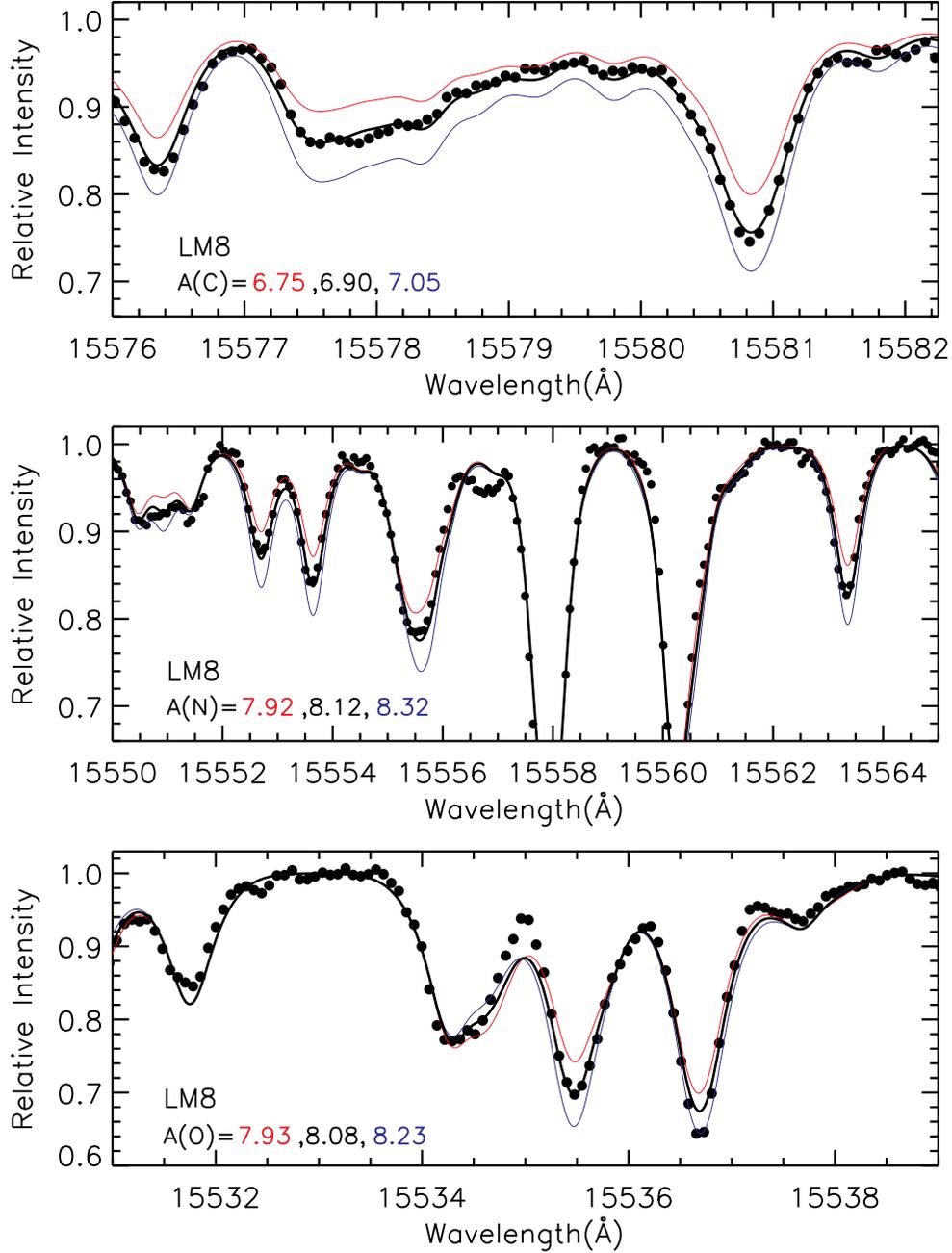}
\caption{Observed spectra (circles) and synthetic spectra 
for C (upper), N (middle), and O (lower) in LM8. The synthetic 
spectra show the best fit (thick black line) and unsatisfactory 
fits (thin red and blue lines) 
$A$(C) $\pm$ 0.15 dex, $A$(N) $\pm$ 0.20 dex, and $A$(O) $\pm$ 0.15 dex. 
\label{fig:cn}}
\end{figure}

\begin{figure}
\epsscale{0.8}
\plotone{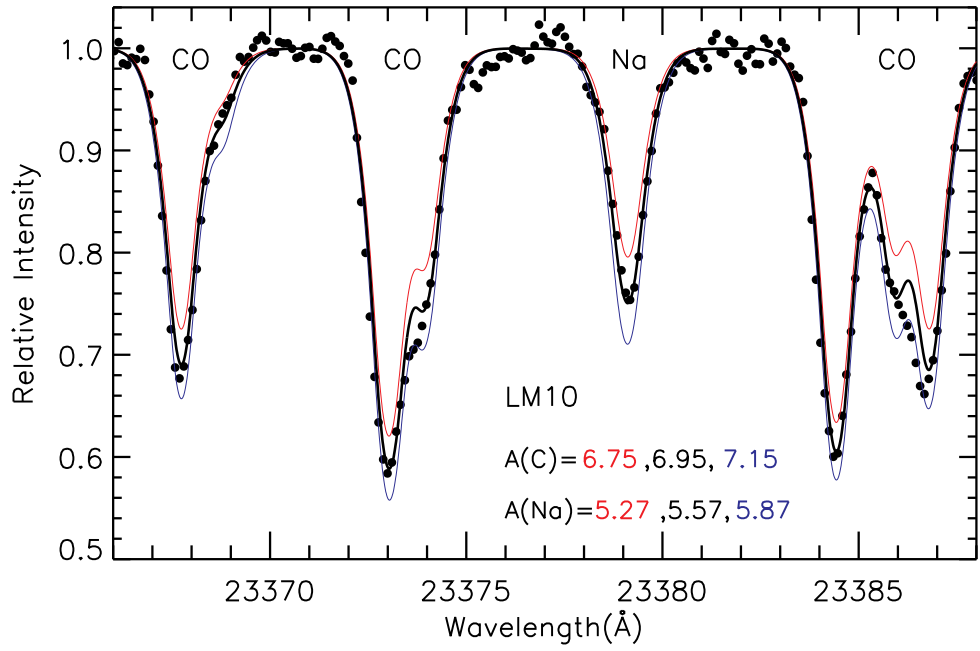}
\caption{Observed spectra (circles) and synthetic spectra 
for LM10. 
The lines show syntheses with different C and Na abundances. The positions of 
CO and Na lines are shown. 
\label{fig:na}}
\end{figure}

\begin{figure}
\epsscale{0.8}
\plotone{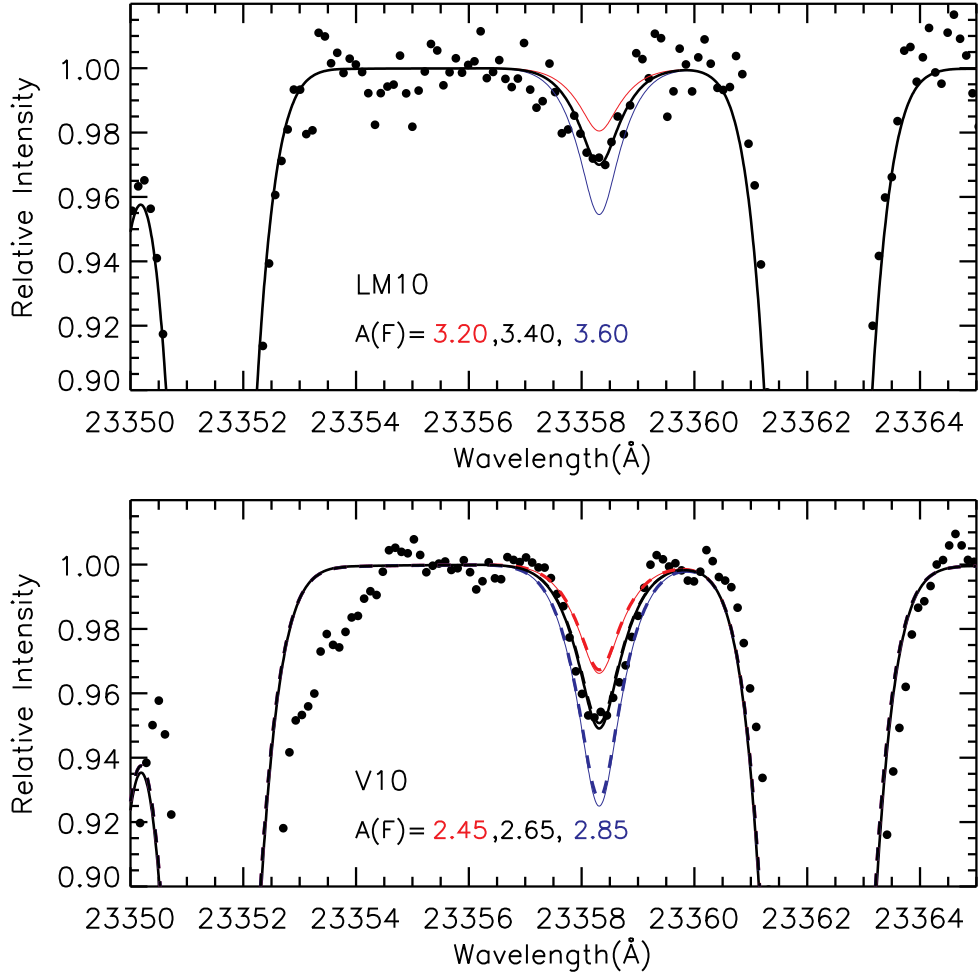}
\caption{Observed spectra (circles) and synthetic spectra 
for F in LM10 (upper) and V10 (lower). 
In the lower panel, 
two sets of syntheses are plotted corresponding to 
$\log g$ = +0.22 (dashed line) and 0.00 (solid line). The lines 
are indistinguishable and for this element in this star, 
the extrapolated abundance 
for $\log g$ = $-$0.22 is $A$(F) = 2.65. 
\label{fig:f}}
\end{figure}

Abundances for a given line were derived by comparing synthetic spectra 
with observed spectra. The synthetic spectra were generated using 
the local thermodynamic equilibrium (LTE) 
stellar line analysis program 
MOOG \citep{moog} and LTE model atmospheres from 
the \citet{kurucz93} grid. 
First we derived abundances for O from the OH molecular lines at 15535.462\AA, 
15536.705\AA, and 15565.880\AA. 
Next, abundances for C were obtained from the CO molecular lines near 
15576\AA\ as well as 
from the large number of CO lines in the K-band spectra. Finally, N 
abundances were derived from the CN molecular lines at 15552.695\AA, 
15553.642\AA, and 15563.355\AA. Since the abundances of C, N, and O 
are coupled, we iterated until self consistent abundances were obtained, 
which always occurred within one iteration. 
Abundances for F were obtained from the HF 
molecular line at 23358.311\AA. Na abundances 
were derived from the Na\,{\sc i} line at 23379.140\AA. Fe abundances 
were obtained from the Fe\,{\sc i} lines at 15534.260\AA\ and 15537.690\AA\ 
as well as the Fe blend near 15551\AA. 
In Figures \ref{fig:cn}, \ref{fig:na}, and \ref{fig:f}, 
we show examples of synthetic spectra fits to derive 
abundances in our sample and in Table \ref{tab:param}, we present the 
final abundances. 
The full line list used in the generation of synthetic spectra 
was taken from \citet{jorissen92}, \citet{melendez99}, and 
\citet{melendez01,melendez03}. 

The model atmosphere grid does not extend 
below $\log g$ = 0.0. For the four stars with surface gravities 
$\log g <$ 0.0, abundances were extrapolated from nearby models, e.g., 
for V10 with $\log g$ = $-$0.22, 
we determined abundances for $\log g$ = +0.22 and $\log g$ = 0.00 
and adopted $A$(X) = $A$(X)$_{{\rm log~}g=0.00}$ + 
($A$(X)$_{{\rm log}~g=0.00}$ $-$ 
 $A$(X)$_{{\rm log}~g=0.22}$). 
We checked the extrapolated results by measuring abundances at an 
additional value of $\log g$. 
For the example above our additional measurement was at 
$\log g$ = +0.44. We note that the derived abundances at 
$\log g$ = +0.44, +0.22, and 0.00 were essentially linear such that 
an extrapolation within the grid provides accurate results. 
Although we are extrapolating beyond the grid, we regard the steps in 
surface gravity as small, $\Delta\log g$ $\le$ 
0.26, and so we anticipate that our results should be reliable. 
The abundance dependences upon the stellar 
parameters are shown in Table \ref{tab:parvar}. 

\section{Results}
\label{sec:results}

Based on optical ($V$ vs.\ $B-V$) and infrared ($K$ vs.\ $J-K$)
color-magnitude diagrams, V8 
is a likely asymptotic giant branch (AGB) star. The radial 
velocity and line strengths are consistent with cluster membership, 
as expected given the proper-motion selection criterion. For this star, the 
lines are considerably broader than in the rest of the sample. 
To match the observed spectra, 
the synthetic spectra for V8 were convolved with a 
Gaussian of width 16 \kms\ which represents the combined effect of 
the instrumental profile (6 \kms), 
atmospheric turbulence, and stellar rotation. 
For the remaining stars, the synthetic spectra were convolved with a 
Gaussian of typical width 10 \kms\ to match the observed spectra. 

For the elements C, N, O, Na, and F, we find large star-to-star 
abundance variations ($\sim$0.6 dex) even within our small sample. 
For these elements, the amplitude of the abundance variation 
far exceeds the measurement uncertainties. 
In this respect, NGC 6712 behaves like all other well studied 
Galactic globular clusters. We also find that the Fe abundance 
does not show any star-to-star abundance variation, although we note 
that the number of Fe lines available in our wavelength regions 
is very small. The dispersion in Fe abundances within our sample 
($\sigma$ = 0.04 dex) 
can be attributed entirely to the measurement uncertainties. 

\begin{figure}
\epsscale{0.8}
\plotone{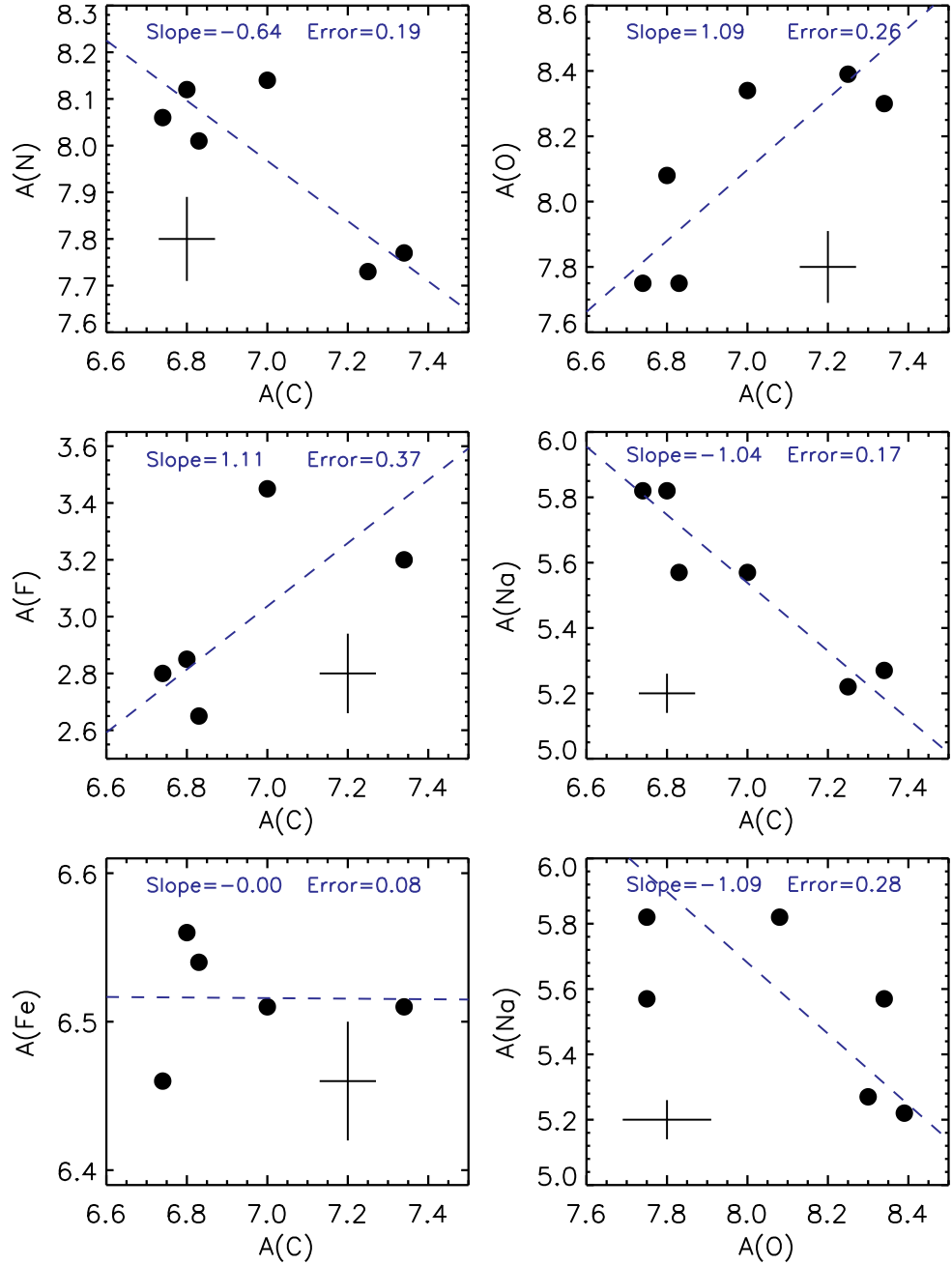}
\caption{Elemental abundances $A$(X) vs.\ $A$(C) as well as 
$A$(Na) vs.\ $A$(O) (lower right panel). A representative 
error bar is shown. The dashed line is the linear least squares 
fit to the data (slope and associated error are included).  
\label{fig:cnofna}}
\end{figure}

In Figure \ref{fig:cnofna}, we plot the abundances of N, O, F, Na, and 
Fe against C as well as O vs.\ Na. 
As seen in all globular clusters, the abundances of 
C and N are anticorrelated and the abundances of C and O are 
correlated. In this figure, we fit a straight line to the data 
taking into account both the $x$ and $y$ errors. We show the formal slope 
of the fitted line as well as the 1-$\sigma$ uncertainty in the slope. 
The C-N anticorrelation is significant at the 3-$\sigma$ level and 
the C-O correlation is significant at the 4-$\sigma$ level. 
We also find that the Na abundances are anticorrelated with C 
at the 6-$\sigma$ level and that 
Na is anticorrelated with O at the 4-$\sigma$ level. 

The F abundance shows a large star-to-star variation. In Figure 
\ref{fig:cnofna}, the F abundances are correlated with C 
at the 3-$\sigma$ level. Therefore, F is also correlated with O and 
anticorrelated with N and Na. The amplitude of the F abundance 
variation ($\Delta$$A$(F) = 0.80 dex) 
exceeds the amplitude of the O variation ($\Delta$$A$(O) = 0.64 dex). 
(Given the measurement uncertainties $\sigma$A(O) = 0.11 dex and 
$\sigma$A(F) = 0.14 dex, O and F may have comparable abundances.) 
Indeed, of the 
elements measured in our sample, F exhibits the largest amplitude 
abundance variation. 

We find that the sum of C+N+O is constant 
in NGC 6712 within the measurement uncertainties 
(C+N+O is not correlated with the C abundance). 
Finally, we note that the Fe abundances are not correlated with C. 
Adopting a solar abundance $A$(Fe)$_\odot$ = 7.48, we find a mean 
cluster abundance [Fe/H] = $-$0.96 $\pm$ 0.02 ($\sigma$ = 0.04) 
which is in good agreement with previous estimates for this cluster 
by \citet{zinn84}, [Fe/H] = $-$1.01 and \citet{jasniewicz04}, 
[Fe/H] = $-$1.2. 

\section{Discussion}
\label{sec:discussion}

\subsection{Abundance comparison between NGC 6712 and M4}

M4 is an ideal globular cluster with which 
to compare the chemical abundances in NGC 6712. M4 is the 
only other cluster in which F abundances 
have been measured in more than two stars 
\citep{smith05}, it has a comparable metallicity 
([Fe/H]$_{\rm M4}$ = $-$1.20 and [Fe/H]$_{\rm NGC~6712}$ = $-$1.01 
[\citealt{harris96}]), and the orbital parameters are very similar 
($R_{\rm apocentric}^{\rm M4}$ = 5.9 $\pm$ 0.3 kpc, 
$R_{\rm pericentric}^{\rm M4}$ = 0.6 $\pm$ 0.1 kpc, and 
$Z_{\rm max}^{\rm M4}$ = 1.5 $\pm$ 0.4 kpc and 
$R_{\rm apocentric}^{\rm NGC~6712}$ = 6.2  $\pm$ 0.3kpc, 
$R_{\rm pericentric}^{\rm NGC~6712}$ = 0.9 $\pm$ 0.1 kpc, and 
$Z_{\rm max}^{\rm NGC~6712}$ = 0.9 $\pm$ 0.2 kpc [\citealt{dinescu99}]). 
(In the globular cluster $\omega$ Cen, F has been measured in one star 
and an upper limit measured in another star [\citealt{cunha03}].) 
However, we note that M4 may be uniquely enriched in $s$-process 
elements among the Galactic globular clusters \citep{M4,pritzl05,rbpbm4m5}, 
with the usual exception of $\omega$ Cen \citep{norris95b,smith00}. 

\begin{figure}
\epsscale{0.8}
\plotone{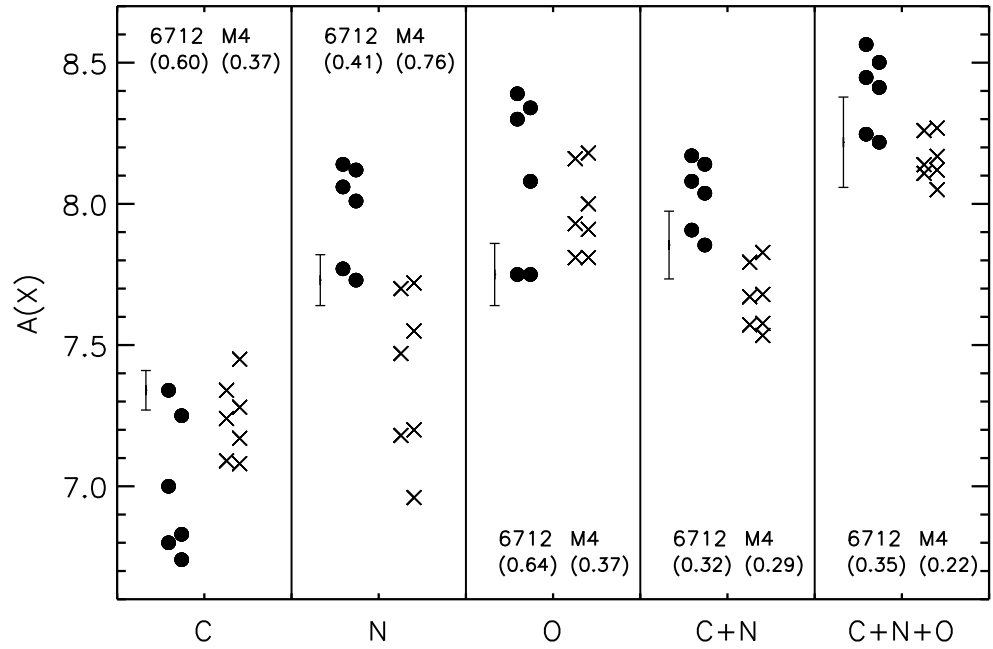}
\caption{The abundance distribution of $A$(C), $A$(N), $A$(O), $A$(CN), 
and $A$(CNO) for NGC 6712 (circles) and M4 (crosses). The M4 data are 
from \citet{smith05}. The amplitude 
of the abundance dispersion is shown along with a representative error 
bar. \label{fig:cno}}
\end{figure}

In Figure \ref{fig:cno} we show the 
abundance ranges for C, N, and O for our six stars in NGC 6712 and the seven 
stars in M4 \citep{smith05}. In both clusters, 
the targets are located near the tip of the red giant branch. 
Within the small samples, NGC 6712 may have slightly larger 
abundance amplitudes for C, O, and C+N+O, the abundance amplitude 
for C+N is very similar for these clusters,  
and M4 has a larger abundance amplitude for N. 
The mean cluster 
abundances are very similar for O. However, NGC 6712 may 
have higher mean abundances of N, C+N, and C+N+O along with a lower  
mean C abundance than M4. 
The higher N abundances and lower C abundances in NGC 6712 
relative to M4 suggests that the extent of CN-cycling may have been 
greater in NGC 6712. However, 
NGC 6712 was formed from gas with higher amounts of C+N and C+N+O  than 
M4. 

\begin{figure}
\epsscale{0.8}
\plotone{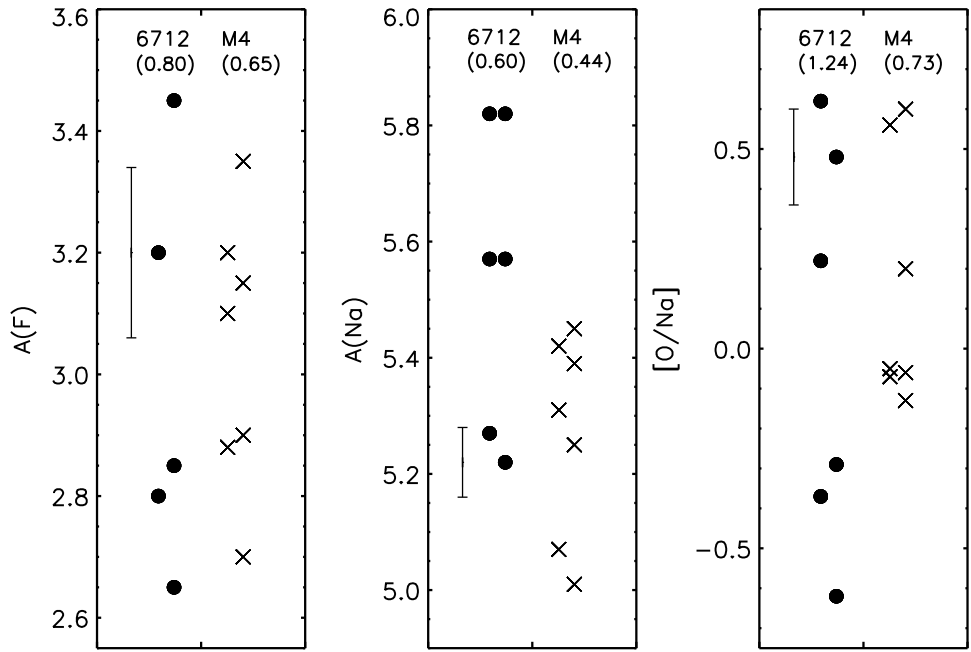}
\caption{Same as Figure \ref{fig:cno} but for $A$(F), $A$(Na), 
and [O/Na]. \label{fig:fnaona}}
\end{figure}

In Figure \ref{fig:fnaona} we show the abundance ranges for F, Na, 
and the ratio [O/Na] for NGC 6712 and M4. 
Within the small samples, NGC 6712 may have  
larger abundance amplitudes for F, Na, and [O/Na] than M4. The mean cluster  
abundances of F are in agreement, however NGC 6712 has a higher mean 
Na abundance and a slightly lower [O/Na] ratio than M4, which is consistent 
with a higher degree of hydrogen burning via ($p$,$\gamma$) reactions. 
Larger samples are required to fully appreciate the abundance 
differences between these two clusters whose orbital parameters 
are very similar. 

\subsection{F destruction and constraints on AGB nucleosynthesis}

The general behavior of the abundances of F 
with respect to C, N, O, and Na in NGC 6712 is identical to that seen 
in M4 \citep{smith05}. 
We reiterate that in both NGC 6712 and M4, the amplitude of the F 
abundance variation is comparable to, or exceeds, 
the amplitude of the O variation. 
Therefore, any scenario invoked to explain the light element abundance 
variations in globular clusters must account for these large F variations. 

In sufficiently massive AGB stars, the 
base of the convective envelope can reach temperatures that permit 
hydrogen burning, a process called hot-bottom burning (HBB) 
\citep{scalo75}. 
Hot-bottom burning can qualitatively produce the required 
C, N, O, Na, Mg, and Al abundance patterns observed in globular 
clusters,  
and intermediate-mass AGB stars have long been suspected of 
producing the light element abundance variations \citep{cottrell81}, 
although AGB models have thus far failed to match the observations
\citep{fenner04,karakas06b}. 
An additional signature of 
HBB is F destruction via $^{19}$F($p$,$\alpha$)$^{16}$O 
\citep{mowlavi96,lugaro04,smith05}. 
In contrast, low mass AGB stars produce F \citep{jorissen92,forestini92}. 

If AGB stars are solely responsible for the F and O abundance 
variations in NGC 6712, theoretical yields 
\citep{karakas03,karakas07,karakas08b}
offer insight into the range of possible masses of these stars. 
These models indicate that F may be destroyed by up to 1 dex 
while O is destroyed by up to 0.5 dex 
during HBB in 5$M_\odot$ and 6$M_\odot$ $Z$ = 0.004 AGB stars, in 
general agreement with the observations. 
However, F destruction ceases and indeed F production begins 
to occur again when HBB is terminated, 
and it is during this phase when much of the mass loss occurs. 
Therefore, even for the most massive stars, the AGB winds contain 
material with O and F depleted by similar amounts. 
The main uncertainties in these models are convection and mass loss. 
Convection determines the efficiency of HBB as well 
as the HBB lifetime \citep{ventura05a,ventura05b}. 
The mass-loss rate determines when the mass is lost from the star. 
For example, a stronger mass-loss rate may result in more mass 
lost when the star was O and F poor but with a larger 
degree of F depletion. 
From 5$M_{\odot}$ models of [Fe/H] $\sim -2.3$ computed for 
\citet{karakas06a}, we estimate that the fluorine yields can vary by 
up to a factor of $\sim$3 by changing the mass-loss rate. 
If massive metal-poor 
AGB stars are responsible for the F and O variations in NGC 6712 and M4, and 
if the F variation exceeds that of O, 
then the models would require both stronger mass loss and more 
efficient convection. 

Results presented by \citet{izzard07}, that use updated NeNa and MgAl 
hydrogen burning rates, gave fluorine abundances increased by a factor of 
72. This last result is net production of $^{19}$F as opposed to 
destruction by HBB in the former models, and serves to illustrate 
just how uncertain the AGB models are to variations in the input physics 
and to the nuclear uncertainties, especially at the range of temperatures 
found in the H and He-burning shells of AGB stars. 
(We refer the reader to the discussions in \citet{lugaro04,lugaro08} and 
\citet{izzard07} and references therein 
for an overview of the current uncertainties regarding AGB model yields 
for fluorine and other light elements.) 
Additional observations of F in AGB stars, such as 
those presented by \citet{uttenthaler08} 
as well as measurements 
in higher mass AGB stars are critical to constrain the AGB models. 

In addition to HBB in intermediate-mass AGB stars, 
another possible source of these abundance anomalies is massive 
stars \citep{prantzos06,smith06,decressin06}. 
While massive stars will also destroy F \citep{prantzos07}, quantitative 
yields for F and O would be of interest to constrain all currently 
proposed sources  
of the globular cluster abundance variations. At present, 
the star-to-star F abundance variations in NGC 6712 and M4 
could be explained by pollution from either a generation of massive 
stars or intermediate-mass AGB stars that underwent hot-bottom burning. 

\subsection{A comparison of [O/Fe] in NGC 6712 with 
the general bulge trend}

The O abundances vary from star-to-star in globular clusters 
(e.g., \citealt{kraft94}). Stars with high O abundances also show high Mg 
abundances along with low Na and Al. We refer to these stars 
as ``normal'' because comparisons have shown that 
the abundance patterns of these 
cluster stars are in accord with field stars at the same metallicity.
At the opposite end of the abundance distribution in globular clusters 
lie the O-poor, Mg-poor, Na-rich, 
and Al-rich stars which we refer to as ``polluted''. No field stars 
have been observed with compositions matching these ``polluted'' cluster 
stars. 
In globular clusters like NGC 6752 \citep{6752} and M13 \citep{sneden04a}, 
the ``normal'' stars with 
the highest O abundances have [O/Fe] ratios in 
agreement with field stars at the same metallicity.  

Following \citet{melendez08}, we adopt solar abundances of 
$A$(O)$_\odot$ = 8.72 and $A$(Fe)$_\odot$ = 7.48 
which are similar to the \citet{asplund05} values based on 
3D hydrodynamical model atmospheres. 
For NGC 6712, our highest relative abundance is [O/Fe] = 0.59. 
We assume that this star is ``normal'' and that this O abundance 
is representative of the initial cluster value prior to the 
processes which produced star-to-star variations in the light element 
abundances. Within the measurement uncertainties, the O abundance for 
NGC 6712 is comparable to the values recently measured in bulge giants  
by \citet{melendez08}, who showed through a homogeneous 
differential analysis that the thick disk and bulge (and halo) had [O/Fe] 
ratios in agreement at a given [Fe/H]. 
Therefore, we tentatively conclude that the [O/Fe] 
ratio in ``normal'' giants in NGC 6712 is in agreement with the 
general bulge trend (and halo stars at the same metallicity).  

\subsection{A comparison of fluorine in NGC 6712 with other 
Galactic populations}

\begin{figure}
\epsscale{0.6}
\plotone{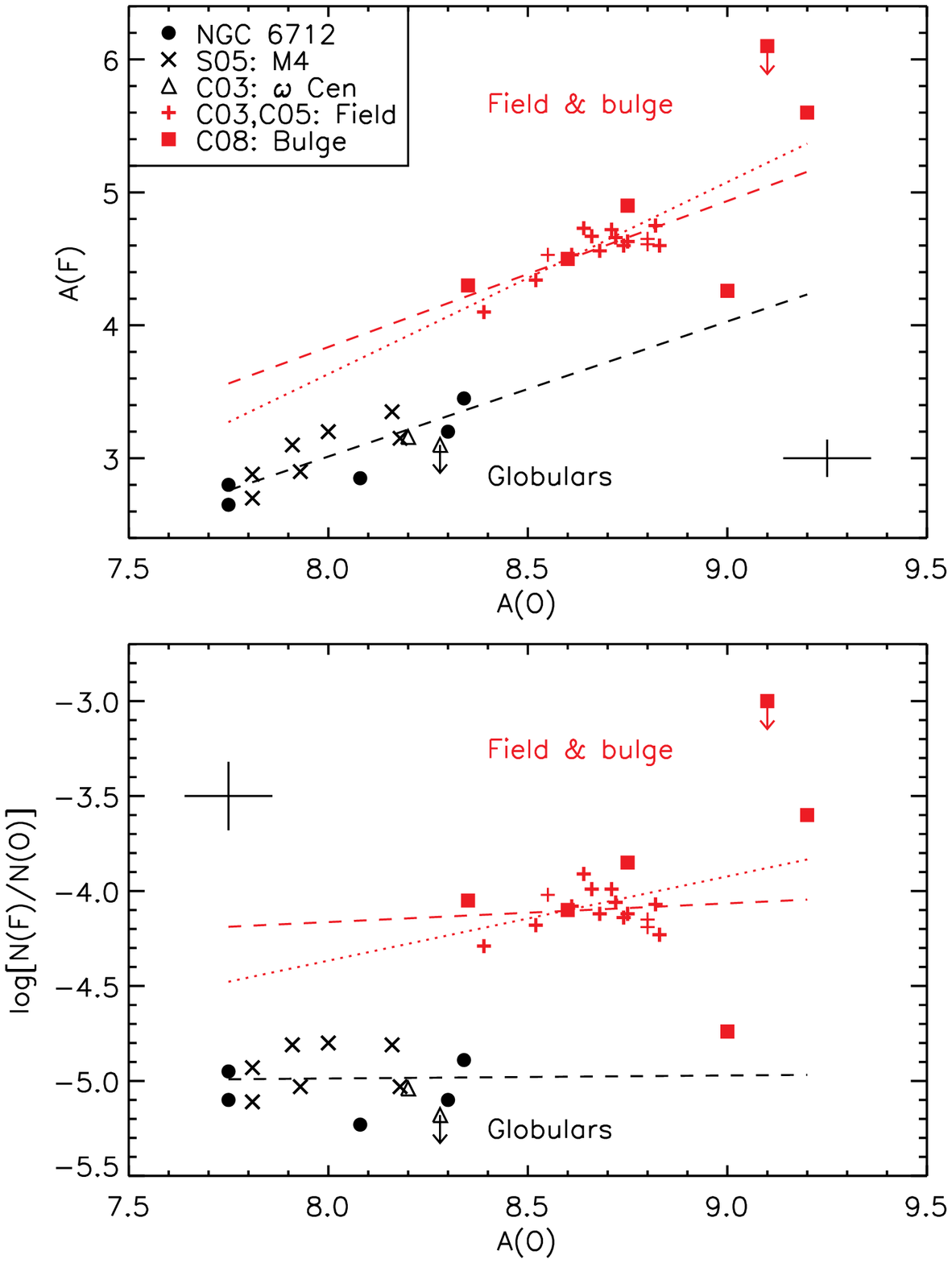}
\caption{$A$(F) vs.\ $A$(O) (upper) and $\log$[N(F)/N(O)] vs.\ 
$A$(O) (lower). NGC 6712 (black circles), 
M4 (black crosses: \citealt{smith05}), 
$\omega$ Cen (black triangles: \citealt{cunha03}), 
bulge stars (red circles: \citealt{cunha08}), and 
field stars (red plus signs: \citealt{cunha03} and \citealt{cunha05}) 
are shown. A representative error bar is shown. 
The red and black 
dashed lines are the linear least squares fits to the field+bulge 
and globular cluster data respectively (excluding upper limits). 
The dotted red line is the fit to the field+bulge data excluding 
the upper limits and the bulge star with $A$(O) = 9.0. \label{fig:fvso}}
\end{figure}

The fluorine abundances measured in NGC 6712 are now compared to those
from other samples of Galactic stars, which include field stars 
\citep{cunha03,cunha05}, along with bulge red giants
\citep{cunha08}, as well as the measurements for the globular
clusters M4 \citep{smith05} and $\omega$ Cen \citep{cunha03}. 
As is the case for oxygen, the
assumption is made that the highest fluorine abundances in NGC 6712 (as well
as for M4) represent the initial cluster value prior to the processes which
produced the globular cluster star-to-star abundance variations. 
In Figure \ref{fig:fvso} 
are plotted the abundances of A(F) versus A(O) (top panel) and
log[N(F)/N(O)] versus A(O) (bottom panel) for all stars from the various
studies.  As a first point of comparison, it is found that the most F-rich
stars in NGC 6712 (and in M4) are underabundant in fluorine when compared
to most of the bulge and field stars that have comparable values of
A(O) $\sim$ 8.2-8.4 (here oxygen is used as a proxy for metallicity). 
Most of the field stars and bulge stars fall along a similar distribution 
in the A(F) versus A(O) diagram, while the globular cluster stars seem to
define a different trend. 

The straight lines shown in the top panel of Figure \ref{fig:fvso} 
represent linear fits
to the globular cluster data and, in addition, fits to the field plus bulge
star points; extrapolations of these lines do not intersect and confirm
the impression that the globular clusters have a distinct mixture of F and
O abundances when compared to the field and bulge stars.  This observation
is based on the stellar samples studied to date, with a still small
metallicity overlap; the behavior of fluorine 
in the field
has not been probed below oxygen abundances A(O) $\sim$ 8.4.
However, to have a single distribution, or curve, of
A(F) to A(O) fit both sets of data (field and globular clusters) 
would require a rapid, nearly
discontinuous drop of about 0.8 dex in the fluorine abundance near an oxygen
abundance of A(O) $\sim$ 8.2-8.4.  
It is also noted that the rather oxygen-rich
bulge giant (BMB 78) studied by \citet{cunha08}, with A(O) = 9.0, but a
low F abundance of A(F) = 4.26, falls along the line extrapolated from the
globular cluster stars. 

Another way to compare F and O abundances is shown in the bottom panel of
Figure \ref{fig:fvso}, 
with the ratio of F/O plotted as a function of the oxygen
abundance.  Here again there appears to be a rather sharp, near
discontinuity in the values of F/O in the globular clusters in comparison
to the field and bulge stars near A(O) $\sim$ 8.3.  In the globular clusters,
the values of F/O remain nearly constant as the oxygen abundance varies
and this is due to the depletion of both $^{16}$O and $^{19}$F by the
H-burning processes that shape the peculiar chemical evolution found in
the globular clusters; inspection of the F and O abundances in both M4 and
NGC 6712 reveals nearly equal decreases in both $^{19}$F and $^{16}$O 
(as expected to occur in only 
the most massive AGB stars, as discussed above), which
results in nearly constant ratios of F/O within the cluster stars.  Note that
Figure 2 in \citet{cunha08} shows predictions for $^{19}$F production
via neutrino nucleosynthesis in SNe II taken from the \citet{woosley95}
models, as well as the approximate downward revisions to the fluorine
yields as suggested by \citet{heger05}.  Neutrino nucleosynthesis
predicts values of log[N(F)/N(O)] $\sim$ $-$5.0 
for models with oxygen abundances
of about A(O) = 8.4, which matches the envelope of values found for the stars
in NGC 6712, M4, as well as the two stars studied to date in $\omega$ Cen.

Within the uncertainties in the measurements, along with the models, it is
suggested that the $^{19}$F observed in the globular clusters could have been
created by neutrino nucleosynthesis alone (with the fluorine arising mostly
from core-collapse neutrinos spalling $^{20}$Ne).  The increased values of
F/O found in the field and bulge stars require additional sources of $^{19}$F,
which have been discussed and modeled by \citet{renda04} and discussed
in \citet{cunha08}, and consist of Wolf-Rayet winds (whose yields of fluorine
increase substantially with stellar envelope metallicity), along with thermally
pulsing AGB stars.  Such a picture would suggest that the
globular clusters are less polluted by Wolf Rayet winds and low-mass AGB
stars (AGB stars with M $>$ 2-3M$_{\odot}$ destroy $^{19}$F) than either the
bulge or field stars with metallicities greater than about one-third solar.
If the globular clusters represent the remnants of systems that formed from
gas that was chemically seeded by very metal-poor SN II, the low values of
F/O represent the ``chemical memory'' of this enrichment.

Although ``normal'' globular cluster stars have compositions that are 
indistinguishable from field stars at the same metallicity, as discussed 
above, $^{19}$F is an exception. Another exception 
is represented by the minor isotopes of Mg, whose ratios 
$^{25}$Mg/$^{24}$Mg and $^{26}$Mg/$^{24}$Mg in ``normal'' cluster stars 
exceed the values found in field stars at the same metallicity 
\citep{shetrone96b,6752,mghdwarf,mghsubaru}. While the contribution, 
or absence, of Wolf Rayet winds and/or low-mass AGB stars provides a 
plausible explanation for the F discrepancy 
as discussed above, the situation for the Mg 
isotopes is less clear. One explanation is that 
the entire globular cluster was polluted by intermediate-mass AGB 
stars which raised the low abundances of $^{25}$Mg and $^{26}$Mg 
provided by supernovae \citep{fenner03} 
to the high levels observed. 

Finally, the bulge star BMB 78, with unusually low F, may be an O-rich star 
whose F abundance could be attributed solely to neutrino nucleosynthesis 
(i.e., the F in this star 
has experienced little or no contribution from Wolf Rayet or AGB stars). 
As discussed by \citet{cunha08}, this star could therefore have important 
implications for the inhomogeneous chemical evolution of the bulge. 
An additional, and highly speculative, explanation for the unusual F and O 
abundances in BMB 78 is that 
this star was born in a globular cluster 
but was subsequently stripped away. 

\subsection{Light element abundance variations and implications for 
Galactic formation}

Star-to-star abundance variations for the light elements have been 
found in every well studied Galactic globular cluster. Such abundance 
patterns are the signature of hydrogen burning at high temperatures. 
The currently favored candidates are intermediate-mass AGB stars 
and massive stars. 
Our abundance measurements in NGC 6712 
provide new and critical information. Specifically, the F abundance 
is found to vary from star-to-star with an amplitude comparable to, 
or possibly exceeding,  
that of O. Therefore, the two globular clusters in which F has been 
measured in more than two stars both show large abundance variations. 

The fact that NGC 6712 exhibits large star-to-star abundance variations 
of the light elements has implications for Galactic 
formation. The current mass in globular clusters is small, but in the past 
there may have been many more clusters. Some fraction of field stars 
may have been born in globular clusters that were subsequently 
destroyed by the 
Galactic tidal field. However, the abundance signature of globular 
clusters, O, F, Na, Mg, and Al variations, has never been 
observed in field stars to date. 
(C and N variations are found in field halo stars
as well as cluster stars [e.g., \citealt{gratton00b}] 
and can be attributed to internal nucleosynthesis 
and mixing with the observed stars.)
Current estimates suggest that globular clusters 
comprise roughly 2\% of the mass of the stellar halo \citep{freeman02}. 
If we arbitrarily assume that one globular cluster was destroyed for every 
surviving cluster (i.e., the initial globular cluster population 
was double the current population), 
then for every 50 field halo stars observed, only 1 star 
would come from a disrupted globular cluster. However, not every 
star in a given globular cluster has peculiar O, F, Na, Mg, and Al abundances 
with respect to field stars at the same metallicity (e.g., see the 
earlier discussion on ``normal'' stars). 
If we arbitrarily assume that half the stars in globular clusters 
have distinct abundances of O to Al relative to field stars at the same 
metallicity, then  100 field halo stars need to observed 
to find one star whose chemical abundances indicate 
that it was born in a globular cluster. Nevertheless, no field halo stars 
have been identified that show large Na and Al enhancements along with  
large O depletions. \citet{gratton00b} investigated a large sample of 
105 stars with $-$2 $\le$ [Fe/H] $\le$ $-$1, and so the non-detection 
of the globular cluster abundance anomalies 
in field stars is not due to a lack of effort (although larger samples 
may be needed). 
NGC 6712 and Pal 5 are tidally disrupted globular 
clusters which have almost certainly contributed stars to the disk and 
halo. Both clusters show large abundance variations for light elements 
which suggests that clusters like NGC 6712 and Pal 5 cannot have provided 
many field stars and/or field stars did not form in environments with 
chemical evolution histories like NGC 6712 and Pal 5. 

\subsection{Constraints upon the initial cluster mass from abundance
variations} 

Based on the present day luminosity function, the high luminosity 
x-ray source, and the large blue straggler population, it is highly 
likely that NGC 6712 was initially considerably more massive 
\citep{demarchi99,andreuzzi01,paltrinieri01}. Indeed, calculations 
by \citet{takahashi00} 
suggest that NGC 6712 may have lost 99\% of its initial mass such 
that it might have been one of the most massive clusters 
that ever formed in the Galaxy, $M_{\rm initial}$ $\sim$ 10$^7$ $M_\odot$. 

For eight well studied globular clusters, \citet{carretta06b} compared  
the interquartile range (IQR) for [O/Fe], [Na/Fe], [O/Na], and other 
abundance ratios with various physical parameters 
and 
found that the amplitude of the abundance variation 
shows a dependence upon cluster mass, as inferred from 
the absolute magnitude. 
Presumably the light element abundance variations in NGC 6712 (and in all 
clusters) originated early in the life of the cluster. Therefore, 
the currently observed abundance variations offer an independent 
estimate of the original mass of NGC 6712. 

Our measured 
values are IQR[O/Na] = 0.85, IQR[O/Fe] = 0.59, and IQR[Na/Fe] = 0.55 
(adopting solar values of 
$A$(O)$_\odot$ = 8.72,  
$A$(Na)$_\odot$ = 6.17, and
$A$(Fe)$_\odot$ = 7.48.) 
Since our sample size is 
small, we may be underestimating (or overestimating) the true IQRs. 
We fit a straight line to the \citet{carretta06b} data and find 
that the IQRs for [O/Na], [O/Fe], and [Na/Fe] in NGC 6712 correspond 
to absolute magnitudes of $-$9.6, $-$12.7, and $-$11.4 respectively. 
While such an analysis is far from robust, inspection of Figures 12 and 
13 in \citet{carretta06b} indicate that NGC 6712 should be a 
very massive cluster based on the IQRs for [O/Fe], [Na/Fe], and [O/Na]. 
The two most massive globular clusters $\omega$ Cen and M54 have 
absolute magnitudes $-$10.29 and $-$10.01 respectively and both clusters 
are regarded as the nuclei of accreted dwarf galaxies. 
Despite our small sample size which may not measure the true IQRs, 
it is likely that 
NGC 6712 was initially one of the most massive clusters in our Galaxy 
as inferred from the large amplitude light element abundance variations. 
Of great 
interest would be the analysis of a larger number of elements in a 
larger sample of stars in NGC 6712 to identify abundance similarities 
with the massive globular cluster $\omega$ Cen. Given the narrow RGB 
sequence \citep{cudworth88}, a star-to-star spread in Fe seems 
unlikely. 

\section{Concluding remarks}
\label{sec:summary}

Based on high resolution infrared spectra, we derive abundances 
of C, N, O, F, Na, and Fe in six giant stars of the tidally 
disrupted globular cluster NGC 6712. For the elements C, N, O, F, 
and Na, we find large star-to-star abundance variations and correlations 
between these elements, a 
characteristic that NGC 6712 shares with every well studied Galactic 
globular cluster. This is only the second cluster in which F 
abundances have been measured in useful numbers of stars 
and both clusters show F variations whose 
amplitude is comparable to, or exceeds, that of O. Within the limited data, 
globular clusters appear to have lower F abundances than field 
and bulge stars at the same metallicity. 
Of great interest would be 
measurements of F in additional stars in $\omega$ Cen and 
other globular clusters as well as in larger samples of field stars,  
with both samples overlapping in metallicity. 
From the amplitude of the O and Na 
abundance variations, we tentatively confirm that NGC 6712 was 
once one of the most massive clusters in our Galaxy. 

NGC 6712 is a tidally disrupted cluster as revealed through its 
highly unusual luminosity function. Pal 5 is another tidally 
disrupted globular cluster. Both NGC 6712 and Pal 5 
have almost certainly contributed stars to the 
disk and halo. Both clusters exhibit large 
star-to-star abundance variations for light elements, a characteristic 
which has yet to be identified 
in field halo stars. Therefore, the light element abundance variations 
detected in NGC 6712 indicate that clusters like NGC 6712 and Pal 5 
have not provided many field stars and/or field stars did not form 
in environments with chemical enrichment histories like NGC 6712 and Pal 5. 
As pointed out by \citet{smith02}, disrupted globular clusters like 
Pal 5 have lost CN-strong, O-poor, Na-rich, Al-rich stars to the halo 
field. But where are these stars? 
Of great interest would be an abundance analysis of stars within the 
tidal tails of Pal 5 as well as a large-scale dedicated search for 
O, Na, and Al abundance anomalies in field halo stars. 

\acknowledgments
This paper is based on observations obtained with the Phoenix infrared
spectrograph, developed and operated by the National Optical Astronomy
Observatory. 
Based on observations obtained at the Gemini Observatory, which is 
operated by the
Association of Universities for Research in Astronomy, Inc., under a 
cooperative agreement
with the NSF on behalf of the Gemini partnership: the National Science 
Foundation (United
States), the Science and Technology Facilities Council (United Kingdom), the
National Research Council (Canada), CONICYT (Chile), the Australian Research 
Council
(Australia), Minist{\' e}rio da Ci\^{e}ncia e Tecnologia (Brazil) and SECYT 
(Argentina), as program GS-2007B-Q-209. 
This research has made use of the SIMBAD database,
operated at CDS, Strasbourg, France and
NASA's Astrophysics Data System. 
DY thanks Gary Da Costa and Ken Freeman for helpful discussions and 
the anonymous referee for helpful comments. 
This research was 
supported in part by NASA through the American Astronomical Society's Small 
Research Grant Program, the Australian Research Council under grants 
DP0663562 and DP0664105, and FCT (project PTDC/CTE-AST/65971/2006).

\clearpage

\begin{deluxetable}{lccccccc} 
\tabletypesize{\footnotesize}
\tablecolumns{8} 
\tablewidth{0pc} 
\tablecaption{Program stars and observations.\label{tab:prog}}
\tablehead{ 
\colhead{Star} &
\colhead{R.A.\ (J2000)} &
\colhead{Dec.\ (J2000)} &
\colhead{$B$\tablenotemark{a}} &
\colhead{$V$\tablenotemark{a}} &
\colhead{$J$\tablenotemark{b}} &
\colhead{$H$\tablenotemark{b}} &
\colhead{$K$\tablenotemark{b}} 
}
\startdata
V10 & 18 52 57.33 & $-$08 41 43.9 & 15.74 & 13.68 & 9.428 & 8.366 & 8.114 \\
V8 & 18 53 05.65 & $-$08 41 12.2 & 15.24 & 13.32 & 9.391 & 8.596 & 8.274 \\
V21 & 18 52 58.79 & $-$08 42 06.1 & 15.62 & 13.45 & 9.536 & 8.516 & 8.283 \\
LM5 & 18 52 59.31 & $-$08 41 34.5 & 15.61 & 13.63 & 9.906 & 9.037 & 8.764 \\
LM8 & 18 53 06.91 & $-$08 40 54.2 & 15.67 & 13.78 & 10.307 & 9.378 & 9.193 \\
LM10 & 18 53 09.38 & $-$08 43 12.8 & 15.44 & 13.60 & 10.265 & 9.416 & 9.217 \\
\enddata

\tablenotetext{a}{$B$ and $V$ magnitudes from \citet{cudworth88}}
\tablenotetext{b}{$J$, $H$, and $K$ magnitudes from 2MASS \citep{2mass}}

\end{deluxetable}

\begin{deluxetable}{lcccccc} 
\tabletypesize{\footnotesize}
\tablecolumns{7} 
\tablewidth{0pc} 
\tablecaption{Log of observations.\label{tab:obs}}
\tablehead{ 
\colhead{Star} &
\colhead{Date} & 
\colhead{Exp.\ time (s)} & 
\colhead{S/N} & 
\colhead{Date} & 
\colhead{Exp.\ time (s)} & 
\colhead{S/N}  \\
\colhead{} & 
\multicolumn{3}{c}{15550\AA} & 
\multicolumn{3}{c}{23330\AA}
}
\startdata
 \noalign{\vskip +0.5ex}
 \multicolumn{7}{c}{NGC 6712} \cr
 \noalign{\vskip  .8ex}%
 \hline
 \noalign{\vskip -2ex}\\
V10 & 2007 07 27 & 2 $\times$ 260 & 180 & 2007 08 28 & 2 $\times$ 375 & 250 \\
 &  &  &  & 2007 08 29 & 2 $\times$ 375 & 190 \\
V8 & 2007 08 18 & 4 $\times$ 325 & 210 & 2007 08 29 & 2 $\times$ 435 & 200 \\
V21 & 2007 08 18 & 2 $\times$ 300 & 170 & 2007 08 29 & 2 $\times$ 440 & 190 \\
LM5 & 2007 08 18 & 2 $\times$ 490 & 150 & 2007 08 29 & 2 $\times$ 690 & 260 \\
LM8 & 2007 08 18 & 2 $\times$ 680 & 200 & 2007 08 29 & 4 $\times$ 525 & 200 \\
LM10 & 2007 08 18 & 2 $\times$ 710 & 180 & 2007 08 29 & 4 $\times$ 540 & 220 \\
\cutinhead{Radial Velocity Standards}
HD203344 & 2007 07 27 & 4 $\times$ 7 & 450 & 2007 08 29 & 2 $\times$ 8 & 250 \\
HD206642 & 2007 08 18 & 8 $\times$ 6 & 300 \\
\enddata

\end{deluxetable}

\begin{deluxetable}{lccccccc} 
\tabletypesize{\footnotesize}
\tablecolumns{8} 
\tablewidth{0pc} 
\tablecaption{Radial velocities (\kms).\label{tab:rv}}
\tablehead{ 
\colhead{Star} &
\colhead{HJD} & 
\colhead{$V_{\rm rad}$} & 
\colhead{$\sigma$} & 
\colhead{HJD} & 
\colhead{$V_{\rm rad}$} & 
\colhead{$\sigma$} & 
\colhead{Mean $V_{\rm rad}$} \\
\colhead{} & 
\multicolumn{3}{c}{15550\AA} & 
\multicolumn{3}{c}{23330\AA} 
}
\startdata
V10 & 2454308.7893 & $-$104.8 & 0.8 & 2454341.5200 & $-$109.8 & 0.8 & $-$107.3 \\
V8 & 2454330.5523 & $-$106.5 & 2.0 & 2454341.5398 & $-$107.4 & 1.3 & $-$106.9 \\
V21 & 2454330.5807 & $-$106.7 & 0.9 & 2454341.5577 & $-$108.2 & 0.4 & $-$107.4 \\
LM5 & 2454330.6036 & $-$112.6 & 1.2 & 2454341.5753 & $-$117.2 & 0.6 & $-$114.9 \\
LM8 & 2454330.6243 & $-$114.2 & 0.7 & 2454341.6055 & $-$115.4 & 0.5 & $-$114.8 \\
LM10 & 2454330.6487 & $-$101.4 & 1.1 & 2454341.6389 & $-$103.3 & 0.5 & $-$102.3 \\
\enddata

\tablecomments{For the radial velocity standards, we adopted 
HD203344 $V_{\rm rad}$ = $-$88.8 \kms\ and 
HD206642 $V_{\rm rad}$ = $-$58.0 \kms\ \citep{rvs}.}

\end{deluxetable}

\begin{deluxetable}{lccccccccccc} 
\tabletypesize{\scriptsize}
\tablecolumns{12} 
\tablewidth{0pc} 
\tablecaption{Stellar parameters and abundances.\label{tab:param}}
\tablehead{ 
\colhead{Star} &
\colhead{\teff~(K)} &
\colhead{log $g$} &
\colhead{$\xi_t$ (\kms)} &
\colhead{$A$(C)} &
\colhead{$A$(N)} &
\colhead{$A$(O)} &
\colhead{$A$(F)} &
\colhead{$A$(Na)} &
\colhead{$A$(Fe)} &
\colhead{$A$(C+N)} &
\colhead{$A$(C+N+O)} 
}
\startdata
V10 & 3595 & $-$0.22 & 2.04 & 6.83 & 8.01 & 7.75 & 2.65 & 5.57 & 6.54 & 8.04 & 8.22 \\
V8 & 3775 & $-$0.26 & 1.94 & 7.25 & 7.73 & 8.39 & \ldots & 5.22 & \ldots & 7.85 & 8.50 \\
V21 & 3708 & $-$0.26 & 1.98 & 6.74 & 8.06 & 7.75 & 2.80 & 5.82 & 6.46 & 8.08 & 8.25 \\
LM5 & 3820 & $-$0.07 & 1.91 & 7.34 & 7.77 & 8.30 & 3.20 & 5.27 & 6.51 & 7.91 & 8.45 \\
LM8 & 3933 & 0.12 & 1.84 & 6.80 & 8.12 & 8.08 & 2.85 & 5.82 & 6.56 & 8.14 & 8.41 \\
LM10 & 4029 & 0.14 & 1.78 & 7.00 & 8.14 & 8.34 & 3.45 & 5.57 & 6.51 & 8.17 & 8.57 \\
\enddata

\tablecomments{$A$(X)= log[$n$(X)/$n$(H)] + 12.}

\end{deluxetable}

\begin{deluxetable}{lcccccc} 
\tabletypesize{\footnotesize}
\tablecolumns{7} 
\tablewidth{0pc} 
\tablecaption{Abundance dependences on model parameters for 
LM5.\label{tab:parvar}}
\tablehead{ 
\colhead{Species} &
\colhead{\teff\ + 50} & 
\colhead{$\log g$ + 0.2} & 
\colhead{$\xi_t$ + 0.2} & 
\colhead{Total\tablenotemark{a}} 
}
\startdata
$A$(C) & 0.04 & 0.05 & 0.03 & 0.07 \\
$A$(N) & 0.04 & 0.08 & 0.03 & 0.09 \\
$A$(O) & 0.09 & $-$0.05 & $-$0.03 & 0.11 \\
$A$(F) & 0.13 & $-$0.05 & 0.01 & 0.14 \\
$A$(Na) & 0.05 & $-$0.02 & $-$0.03 & 0.06 \\
$A$(Fe) & $-$0.03 & 0.02 & $-$0.02 & 0.04 \\
\enddata

\tablenotetext{a}{The total value is the quadrature sum of the three 
individual abundance dependences}

\end{deluxetable}

\clearpage

\clearpage

\clearpage

\clearpage

\clearpage

\clearpage

\clearpage

\clearpage

\end{document}